# Monte Carlo Study of a $^{137}$Cs calibration field of the China institute of atomic energy

GAO Fei $^{a,b,*}$

$^a$ Radiation Metrology Division, China Institute of Atomic Energy, Beijing, 102413, China

$^b$ National Key Laboratory for Metrology and Calibration Techniques, Beijing, 102413, China

**Abstract** The MCNP code was used to study the characteristics of gamma radiation field with collimated beam geometry. A close-to-reality simulation model of the facility was used for calculation air-kerma along the whole range of source-detector-distance (SDD) along the central beam and air-kerma off-axis beam profiles at two different source-detector-distance (SDD). The simulation results were tested by the measured results which were acquired in the Radiation Metrology Center of CIAE. Other characteristics such as the individual contributions of photons scattered in collimator, floor, walls, mobile platform and other parts of the irradiation halls to the total air kerma rate on the beam axis were calculated for the purpose of future improvement of metrological parameters in CIAE. Finally, factors which influence the simulation results were investigated, including e.g., detector volume effects or source density effects.

## 1 Introduction

Radiation Metrology Division of CIAE obtains many photon sources with different activity values for calibration, metrological research and design new dosimeters for detection and measurement of photon-related quantities. With these sources e.g. cesium-137, colbalt-60 and americium-241, it is possible to obtain air kerma rate from 0.2μGy /h up to 54Gy/h sufficient for all of commonly used dosimeters.

Irradiation field shapes and scattered photons are the most important criterions for irradiation halls. So far, the longitudinal and lateral field shapes in air kerma rate were measured by secondary ionization chambers.

The development of computer technology made it possible to use sophisticated mathematical models to calculate relative parameters of the irradiation fields. In this study, a MCNP-4C Monte Carlo code model of cesium-137 irradiation hall was used for beam profiles verification and for the assessment and analysis of scattered air kerma, which is very important for future improvement of metrological parameters of photon beams in CIAE .

## 2 Experiments

For the purpose of routine measurements, the longitudinal profiles in air kerma rate at distances from 1m to 6.5m from the $^{137}$Cs source were measured with a spherical ionization chamber PTW 32002. The transversal beam profiles were measured in a horizontal line perpendicular to the beam axis at two distances of 1m and 2m from the $^{137}$Cs source. For these measurements，an ionization chamber NE 30c. c. (Nuclear Enterprises ) was used due to its smaller volume (outer diameter is about 4cm ).

$^*$Corresponding author:gaofei@ciae.ac.cn

## 3 Simulations

The source model consist of an active $^{137}$Cs volume, a steel and aluminum capsule, and an aluminum holder. The encapsuled source was positioned inside a lead shielding facility with wall thicknesses up to 20cm. The shielding closely surrounding the source on all sides except for a collimator pyramid, the lead ring–collimator of 30cm in length with an opening angle of 16° was integrated in the lead shielding, as shown in Figure 1.

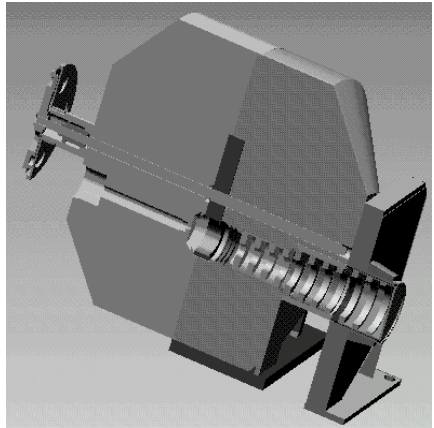

**Figure 1. A simulation model of the lead shielding facility including the lead ring-collimator (right ) and scatter hall (middle ) used for reducing scatter photons of the facility .**

Around the lead shielding facility, a number of concrete and other structures are included, i.e., walls, blocks, ceiling and the mobile platform. The concrete floor of 0.5m thickness, and the air environment are also part of the simulation model, as shown in figure 2.

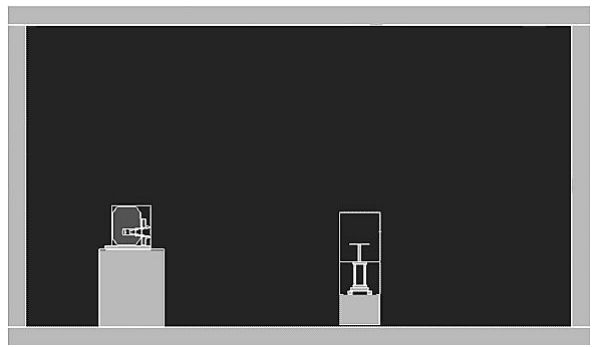

**Figure 2. Side view of the MCNP calculation model showing the lead shielding facility, mobile platform, walls, floor, ceiling and the air environment.**

Four different kinds of surface were necessary for defining the calculation models (number of the used surface in brackets): planes(109),cylinders(34), spheres(1) and cones(2). Seven different materials, elements and compositions are used(mass densities in units of g/cm$^3$ in brackets): lead (11.3), air(0.001293), aluminum(2.78), source(3.988), steel(7.85), collimator(18) and concrete (2.35).

The measurement quality is air-kerma free in air. In the Monte Carlo simulation, air kerma was calculated by folding the simulated photon fluence energy distribution by log-log interplotaed the air-kerma conversion coefficients provided by ICRU[4]. The track-length-estimator tally F4 in MCNP was used to calculate the fluence of photons, the fluence is normalised per emitted source photon. Just as the experiments, longitudinal profiles in air-kerma rate at distances from 1m to 6.5m from the $^{137}$Cs source were simulated. In the Monte Carlo calculations, fluence is detected in spherical air volumes along the central beam axis from 1m to 6.5m in 0.5m interval. Owing to statistical reasons, the diameters of air detector spheres have to be increased from 19cm to 32cm for increasing source-detector distance. The transversal beam profiles were also simulated in a horizontal line perpendicular to the beam axis at two distances of 1m and 2m from the $^{137}$Cs source, the cubic air detectors were used for its lager sensitive volume compared to spherical detectors and columned detectors among the same intervals, as shown in figure 3. As the activity value of the $^{137}$Cs source is not available exactly, only relative air-kerma results are compared in this study.

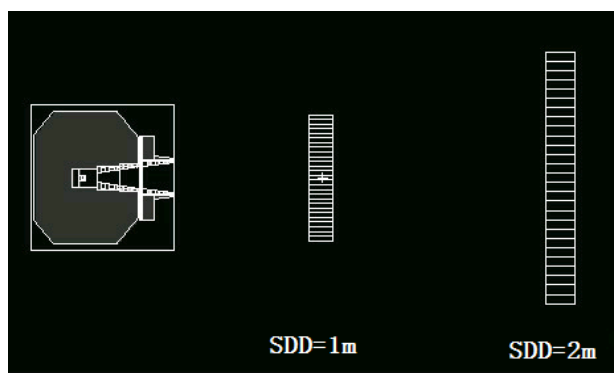

**Figure 3. Top view of Monte Carlo calculation model showing the lead shielding facility(left), cubic air detectors(middle and right) at two different source-detector-distance (SDD) of 1m and 2m.**

## 4.Results and discussion
### 4.1 Lateral field shapes

Fig.4 shows the calculated transversal profile in air-kerma, the results are compared with the experiments. The computed field sizes, shapes and penumbra are in very good agreement with the measurements. Difference between simulated and measured results is within 0.3% on the plateau region and less than 1% on other regions. Geometrical field sizes(the angle of the collimator is 16°) are 22cm(SDD=1m) and 50cm(SDD=2m) respectively. Measured $^{137}$Cs dosimetric field sizes (50% dose level) is ±17.6cm (simulated ±17.3cm ) and ±35.1cm (simulated ±35.0 cm) for 1m and 2m SDD, respectively.

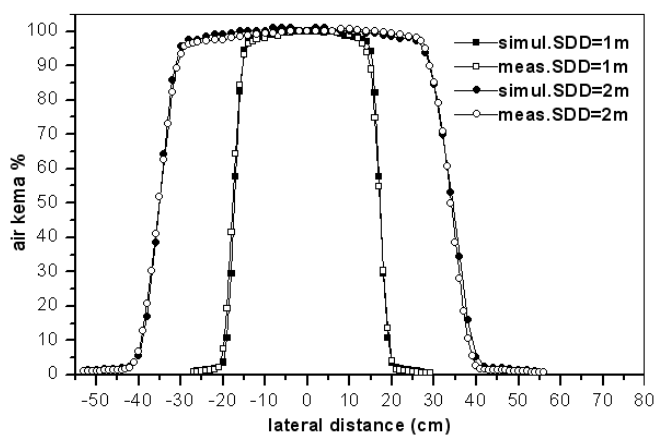

**Fig . 4. Normalized air-kerma off-axis beam profiles at two different source-detector-distance (SDD ) of 1m and 2m.**

### 4.2 Longitudinal dependences

The second comparison between measurement and simulation is along the main beam axis at SDDs in the range of 1m to 6.5m. Figure 5 shows the comparison of PTW Unidos-measured relative air kerma in Cs-137 calibration field with the simulated results in air and in vacuum environment .

Air kerma is multiplied by the square of SDD, the simulated and the measured results are normalized to the 2.5m distance as shown in fig.5, for allowing a better comparison. Comparing with constant values for ideal point sources in vacuum without any scattering facility components, the decreasing curves is simulated in air environment, whereas a increasing curve is simulated in vacuum.

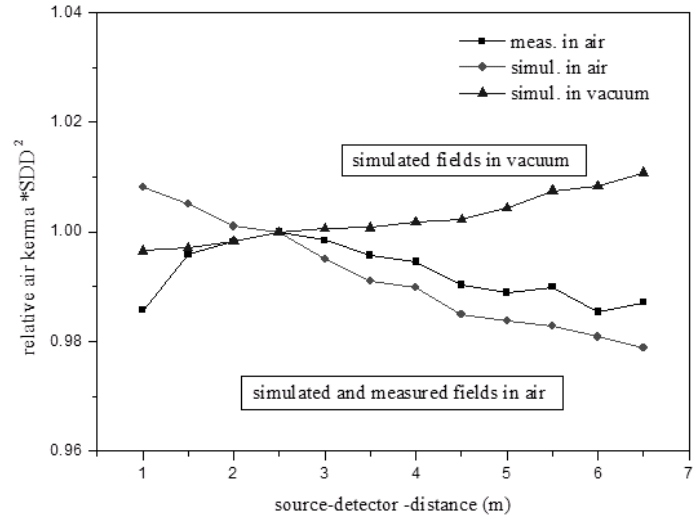

**Fig . 5. Air kerma multiplied by SDD$^2$ along the main beam axis of the Cs-137 calibration field.**

A increasing direction before 2.5m, decreasing direction after 2.5m curve is measured. The increasing is expected due to the contribution of air kerma from another source located in the same radiation facility, whereas the decreasing is expected due to air attenuation. The maximum deviation between simulated and measured air kerma of 2.5% occurs at 1m, less than 1% from 2.5m to 6.5m. A liner increase simulated curve with a increasing rate of 0.3% per m in vacuum is expected due to scattering from various facility components.

**4.3 Facility scattering contribution**

In this paper, the computer transport code MCNP is used to simulate the air kerma which is scattered by facility components of the calibration field. MCNP Monte Carlo code allows distinguishing between photons scattered from various facility components by using the so-called cell-flagging technique. A maximum contribution of 5% from scattered photons by the environment (outside of the source capsule) to the total measured air kerma rate in reference nuclide calibration field is recommended by the ISO standard 4037-1[3].

Figure 6 summarizes the scattered air kerma of all individual parts to the total air kerma at the distance from 2m to 6.5m. The total scattered contribution is increasing from 3.2% to 4.0% (fulfills the ISO 4037-1requirement of a maximum scatter of 5%) toward the end of the calibration hall. The lead collimator with nine apertures (total thickness is 30cm) contributing most to the scattered air kerma, the maximum simulated contribution of the collimator is about 3.3% at the source-detector-distance of 1m, i.e.,63cm from the exit surface of the collimator and the minimum is about 2.6% at the source-detector-distance of 6.5m. At the distance of 6.5m, raising contributions of mobile platform and walls can be seen, as the beam broadens, more photons scattered on walls and mobile platform are detected, the contribution of them up to 0.6% and 0.5%. Scatter contributions from the source holder and concrete floor are less than 0.1%, can be neglected.

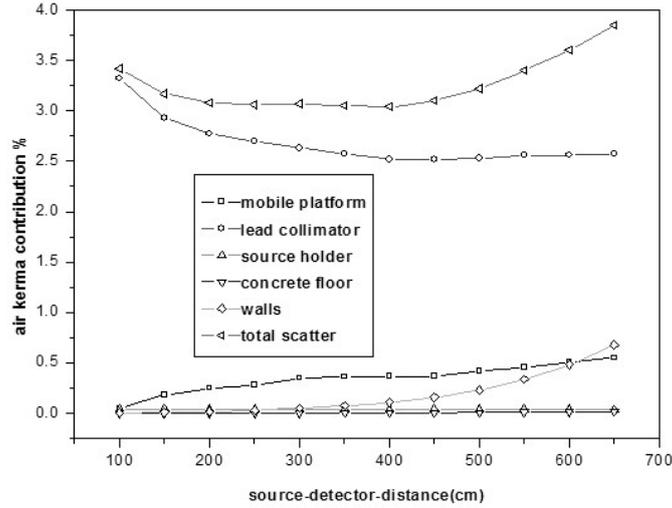

**Fig.6 Relative air-kerma contribution of main calibration facility components to the total air-kerma depending on the source-detector-distance. Main components are collimator(circles ), mobile platform (squares), walls (diamonds), source holder (triangles) and concrete floor(inverse triangles).**

Further measurement should be done to give more information about the scattered air kerma of the calibration hall.

## 5. Uncertainty

The combined standard uncertainty for the measured air-kerma rate is estimated as 2.3% ($k=1$) at the distanceof 6.5m. Table 1 summarises the estimated Monte Carlo simulation uncertainty contributions for the air-kerma rate results. The combined uncertainty for the simulated air-kerma results are all bellow 1.5% (k=1).

**Table 1. List of influence quantities of the simulated air-kerma results and the related uncertainty estimates**

| Influence parameter | Uncertainty estimates (%) |
|---|---|
| Statistical variation | 0.5 |
| Detector volume | 0.6 |
| Estimated source density | 0.1 |
| Finite transport cut-off energy | 0.4 |
| Air-kerma conversion coefficients | 0.1 |
| Cross-sections and physics models | 1.2 |
| Combined uncertainty ($k=1$) | 1.5 |

Due to statistical reasons, the diameters of the detector spheres are chosen between 19cm and 32cm for increasing distance, thus the uncertainty due to deviations from the ideal point detector was roughly estimated 0.6%. The statistical uncertainty estimation is related to the final simulation running number of simulated source photons. The standard deviation due to the random sampling Monte Carlo method is in the range of 0.1% to 0.5% for all SDD as the simulated running number is $1 \cdot 10^9$. The average density of the active material (CsCl column) is about 3.99g /cm$^3$, the estimated uncertainty is 0.1% .

The photon transport energy cut-off order was used in the simulation to accelerate calculating rate, the uncertainty was roughly estimated as 0.4%. The influence of the applied fluence to air-kerma conversion coefficients was estimated by comparing two published datasets. The estimation for the uncertainty of the photon cross-section data and interaction physics in the MCNP code was based on published information in the open literature [5].

**6. Conclusions**

    In this study, the Monte Carlo code MCNP 4C was used to simulate the scatter air-kerma of Cs-137 source in the calibration hall of China Institute of Atomic Energy. The simulation models were tested by the experiments of longitudinal and transveral beam profiles in air kerma. Good agreement between the simulation and the ionization chamber measurements has been shown, but the observed deviations in the simulated results show the need to improve the simulation model. Furthermore, experiment of the scattered photons to the total air kerma on the beam axis should be carried out. The results presented in this paper confirmed the models' accuracy, and hence, these models can be used to learn more about the scattered informations of the calibration hall in the future.